\documentclass[%
 reprint,
 amsmath,amssymb,
 aps,
pre,
]{revtex4-1}

\usepackage{graphicx} 
\usepackage{bm}
\usepackage{txfonts}
\usepackage{mathrsfs}
\usepackage{amsmath}
\usepackage{xcolor}

\usepackage{subcaption}

\def\b0{{\mathbf 0}}

\def\b0{{\mathbf 0}}

\def\eps{\epsilon}

\newcommand{\vk}{\textbf{k}}
\newcommand{\vkd}{\textbf{k}_d}
\newcommand{\vR}{\textbf{R}}

\newcommand{\di}{\mathrm{d}}
\newcommand{\aeps}{\tilde{\epsilon}_{\vk}}

\newcommand{\epsi}{\frac{1}{\psi}}

\def\beq{\begin{equation}}
\def\eeq{\end{equation}}

\begin{document}

\title{Dimensional crossovers and Casimir forces for the Bose gas in anisotropic optical lattices} 
\author{Maciej {\L}ebek}
\affiliation{Institute of Theoretical Physics, Faculty of Physics, University of Warsaw, Pasteura 5, 02-093 Warsaw, Poland}
%
\author{Pawe{\l} Jakubczyk}
\affiliation{Institute of Theoretical Physics, Faculty of Physics, University of Warsaw, Pasteura 5, 02-093 Warsaw, Poland} 
\date{\today}
\begin{abstract}
We consider the Bose gas on a $d$-dimensional anisotropic lattice employing the imperfect (mean-field) gas as a prototype example. We study the dimensional crossover arising as a result of varying the dispersion relation at finite temperature $T$. We analyze in particular situations where one of the relevant effective dimensionalities is located at or below the lower critical dimension, so that the Bose-Einstein condensate becomes expelled from the system by anisotropically modifying the lattice parameters controlling the kinetic term in the Hamiltonian. We clarify the mechanism governing this phenomenon. Subsequently we study the thermodynamic Casimir effect occurring in this system. We compute the exact profile of the scaling function for the Casimir energy. As an effect of strongly anisotropic scale invariance, the Casimir force below or at the critical temperature $T_c$ may be repulsive even for periodic 
boundary conditions. The corresponding Casimir amplitude is universal only in a restricted sense, and the power law governing the decay of the Casimir interaction becomes modified. We also demonstrate that, under certain circumstances, the scaling function is constant for sufficiently large values of the scaling variable, and in consequence is not an analytical function.
At $T>T_c$ the Casimir-like interactions reflect the structure of the correlation function, and, for certain orientations of the confining walls, show exponentially damped oscillatory behavior so that the corresponding force is attractive or repulsive depending on the distance.


\end{abstract}


\maketitle


\section{Introduction}
Ultracold atomic gases in optical lattices have remained a topic of great interest over the last years from both theoretical and experimental points of view.\cite{bloch_many-body_2008, Goldman_2014, Dutta_2015, Baier_2016, Krutitsky_2016} The progressing experimental developments allowed for exploiting physical situations inaccessible in traditional condensed matter setups (and also in continuum cold gases) and stimulated enormous theoretical developments worldwide. 

In this paper we investigate the physics emergent in optical-lattice Bose systems as a result of introducing strong spatial anisotropies giving rise to the presence of (at least) two distinct lengths scales $\xi_{\parallel}$ and $\xi_{\perp}$ controlling the decay of the correlation function in different directions. Both $\xi_{\parallel}$ and $\xi_{\perp}$ diverge at the transition to the condensed phase and are related by the anisotropy exponent $\theta_A$, such that $\xi_{\perp}\sim\xi_{\parallel}^{\theta_A}$. 

First we analyze the possibility of controlling  the Bose-Einstein condensation by anisotropically varying the hopping parameters (or, equivalently,  the dispersion relation). This constitutes an interesting route for inducing the transition between the Bose-Einstein condensed and normal phases, which should be possible in experimentally realized optical lattice systems.   
By suitably tuning the lattice parameters one induces crossovers to physical situations characterized by fractional effective dimensionalities.\cite{Jakubczyk_2018} This may lead in particular to configurations where some of the effective dimensionalities relevant for the system are at or below the lower critical dimension $d_l$, while others are above. This yields certain features of the phase diagram (in particular the crossover scales) not obvious and presumably sometimes hard to access within numerical approaches. Unlike the case, where the transition is tuned by temperature or density, the proposed setup allows for realizing a rich spectrum of universality classes. 
 
 Due to the mean-field character of the studied system, the present analysis may be carried out exactly. Nonetheless, we argue that  many of the studied features are  not necessarily restricted to mean-field models and could also be found in systems characterized by realistic interactions. We believe the analyzed setup might be conceivable in future experiments in optical lattices, to which our results might apply both on the qualitative and 
 quantitative level.
 
The second, separate issue of the present paper concerns the thermodynamic Casimir effect\cite{Mostepanenko_1988, Krech_book, Kardar_1999, Brankov_book, Maciolek_2018}  in anisotropic Bose systems, where the anisotropy is inherited from the lattice. As was indicated in a relatively recent work on the $O(N)$ models in the vicinity of the Lifshitz point,\cite{Burgsmuller_2010} strong anisotropy, manifested by nontrivial scaling of two correlation lengths, leads to the remarkable effect of modifying the power law governing the decay of the Casimir force. In a standard situation the Casimir decay exponent $\zeta_0$ is fully determined by the system dimensionality $d$ and the magnitude of the Casimir interaction is controlled by a universal scaling function, which, strictly at the transition, takes the value of the critical Casimir amplitude. However, according to the prediction of Ref.~\onlinecite{Burgsmuller_2010}, in strongly anisotropic systems, the decay exponent is not completely determined by $d$ and depends \textsl{inter alia}  on the anisotropy exponent $\theta_A$. This is interesting, because, by dimensional analysis, the scaling function must then contain a dimensionful factor, which may originate only from microscopic parameters, thus restricting the universal nature of the scaling function.  
As turns out, the asymptotic expression for the Casimir energy (at or below the critical temperature $T_c$ for condensation)  indeed contains at least one length scale in addition to the system extension $D$ and the scaling function is universal only after the appropriate dimensionful (nonuniversal) coefficient is correctly identified and factored out. 

An equally remarkable result of Ref.~\onlinecite{Burgsmuller_2010} states that for certain orientations of the system relative to the confining walls, the Casimir force is repulsive even for the periodic boundary conditions. We confirm this picture within the present exact study of the imperfect Bose gas. There exist exact statements\cite{Li_1997, Kenneth_2006}, usually formulated in the context of the electrodynamic Casimir effect, concerning the attractive nature of the Casimir force with (\textsl{inter alia}) periodic boundary conditions. The corresponding proofs, however, use the specific form of the inverse propagator (which is quadratic in momentum) and do not carry over to the situations analyzed here.   

 If the temperature is fixed above $T_c$, the Casimir force is exponentially suppressed at distances larger than the correlation length. We find however, that its typically attractive character may be significantly modified by varying the orientation of the confining walls. The Casimir force then shows damped oscillatory behavior and its actual sign depends on the distance $D$ between the confining walls. The obtained behavior should be of relevance for the entire universality class of anisotropic $O(N)$-symmetric models in the limit $N\to\infty$.  

The outline of the paper is as follows: In Sec.~II we give a summary of our results, distilling the conclusions, which, (by virtue of universality) should be independent of the specific microscopic realization. In the following part of the paper (Sec.~III-V) we present a study of the problem employing a concrete microscopic model. In Sec.~III we introduce the imperfect Bose gas on an anisotropic lattice. Its bulk properties in the relevant regime of low temperatures are reviewed with particular focus on the effects caused by anisotropies. In Sec.~IV we present our results on the dimensional crossovers with emphasis on the possibility of tuning the system continuously to a state characterized by the effective dimensionality at or below the lower critical dimension $d_l$ (Sec.~IV B). We give arguments suggesting that the results of this section are not necessarily restricted to mean-field models and may apply to a broad class of systems characterized by realistic microscopic interactions. Sec.~V is independent of Sec.~IV. Here we present our derivation of the expression for the Casimir energy and extract its asymptotic behavior for large separations between the confining walls. We compute and discuss the scaling function for the Casimir energy.  The entire study is carried out by means of an exact analysis. Sec.~VI contains a summary and outlook.    

\section{Statement of the problem and key results}
A canonical equilibrium many-body problem involves a microscopic Hamiltonian $\hat{H}$ comprising a kinetic term and interaction 
\beq
\hat{H}=\hat{H}_{kin}+\hat{V}\;,
\eeq
where, typically, $\hat{H}_{kin}$ is a one-body operator characterized by a dispersion relation $\epsilon_{\bf k}$, which  is quadratic in momentum ${\bf k}$ for $|{\bf k}|$ small. In the vicinity of a continuous  phase transition certain  aspects of the system are universal (i.e. sensitive only to crude characteristics of $\hat{H}$). They however do depend on the asymptotic behavior of $\epsilon_{\bf k}$ at vanishing $|{\bf k}|$, hereafter denoted as $\tilde{\eps_{\textbf{k}} }$. In a lattice system it is possible to engineer the hopping amplitudes so that the dispersion decays more quickly (for example as $|{\bf k}|^4$) if ${\bf k}$ is chosen along a particular direction. In general, the asymptotic expression for the dispersion (at $|{\bf k}|$ small) may be written as 
\beq
 \label{tildeeps2}
 \tilde{\eps_{\textbf{k}} }=\sum_{i=1}^{d} t_i |k_i|^{\alpha_i}, \hspace{1cm} t_i, \, \alpha _i >0\;.
 \eeq
For the specific case of the hypercubic lattice, the exponents $\alpha _i >0$ are natural even numbers. The critical properties then depend on the set $\{\alpha_i\}|_{i=1}^d$, which determines an effective dimensionality $d_{eff}$ of the system. The upper and lower critical dimension of the anisotropic system as well as the critical indices at Bose-Einstein condensation are then the same as those corresponding to the isotropic model in dimensionality $d_{eff}$. We believe that this equivalence (demonstrated for a specific microscopic model) holds true for the $O(N\to\infty)$ universality class, but not beyond. The situation is different for the universal asymptotic shape of the critical line in the phase diagram in the limit of low temperatures (i.e. approaching the quantum critical point) described by the shift exponent $\psi$. Concerning this aspect we argue that the abovementioned correspondence  may possibly be extended to the entire family of universality classes described by the $O(N)$ models in $d$ dimensions. 

By extending the dispersion of Eq.~(\ref{tildeeps2}) accounting for the subleading terms and manipulating the hopping parameters, one may drive the system across physical situations characterized by different effective dimensionalities. We have performed a detailed analysis of this phenomenon in Sec.~IV. Even though the calculation is carried within an exactly soluble mean-field model, we argue that the key aspects related to the $T_c$ phase boundary would be the same for systems with realistic short-ranged interactions.   

A separate problem addressed in Sec.~V concerns the impact of anisotropies inherited from the lattice on the thermodynamic Casimir effect. Our calculation is again carried out within the framework of the imperfect Bose gas model, however, by virtue of universality, the results should be of relevance to the entire $O(N\to\infty)$ universality class in $d$ dimensions with periodic boundary conditions. Here we focus on a hypercubic lattice (wth lattice constant $A$) characterized by the dispersion with asymptotics given by
\begin{equation} 
\label{asssym}
\aeps = \sum_{i=1}^{d-m} t_0 (k_i A)^2 + \sum _{i=d-m+1}^{d} t(k_i A)^4\;,
\end{equation} 
i.e. with a quartic dependence on momentum along $m$ ($1\leq m\leq d$) directions (hereafter refereed to as 'special directions'), and a quadratic form in the remaining $d-m$  directions (hereafter refereed to as 'normal  directions').  
Some of our results may be compared to the study of Ref.~\onlinecite{Burgsmuller_2010}, which rested upon the framework of classical field-theoretic approach built with the vicinity of the Lifshitz point in mind. An important prediction of that study concerns the decay exponent for the Casimir energy and yields
\beq
\label{Diehl_zeta_perp}
\zeta_m=\frac{d-m}{\theta_A}+m-1\;,
\eeq 
for the case where the planar confining walls are perpendicular to one of the special directions, and   
\beq
\label{Diehl_zeta_par}
\zeta_m'=d-m(1-\theta_A)-1
\eeq
for the case of confining walls perpendicular to one of the normal directions. The exponent $\zeta_m$ (or $\zeta_m'$)  replaces the standard value $\zeta_0=d-1$ of the Casimir energy decay exponent. Ref.~\onlinecite{Burgsmuller_2010} also makes a statement concerning the sign of the Casimir interaction predicting the possibility of obtaining a repulsive force. Our results, obtained from an exact calculation departing from a microscopic model are fully in line with Eqs.~(\ref{Diehl_zeta_perp}) and (\ref{Diehl_zeta_par}). In the low-temperature phase, the obtained force is repulsive if the walls are perpendicular to one of the special directions, and attractive otherwise. We extracted the corresponding scaling functions for the Casimir energy, which are universal only after factoring out a dimensionful (model-specific) quantity. These scaling functions should characterize the entire $O(N\to\infty)$ universality class with periodic boundary conditions.  

In the high-temperature phase the correlation lengths are finite and the Casimir energy is exponentially suppressed at large distances $D$. The character of its decay as function of $D$ turns out to depend on the orientation of the walls. For a configuration with walls perpendicular to one of the special directions, the interaction is characterized by (damped) oscillations, so that its sign depends on the distance. In contrast, if the walls are perpendicular to one of the normal directions, the decay is monotonous and the force is always attractive. This feature is a reflection of the properties of the pair correlation function in the high-temperature phase and should not be influenced by the type of symmetry-breaking [whether $O(N\to\infty)$ or $O(2)$].

The above section summarizes the scope of the paper and outlines the key conclusions. The following part (Sec.~III-V) constitutes an analysis carried our within the framework of a lattice variant of the imperfect (mean-field) Bose gas.

\section{Model and its bulk solution}
We consider bosons on a lattice at a fixed temperature $T$, chemical potential $\mu$ and contained within the volume $V=L^d$. The system is governed by the Hamiltonian 
\begin{equation}
 \hat{H}=\sum_{\bf k} \epsilon_{\bf k}\hat{n}_{\bf k}+\frac{a}{2V}\hat{N}^2 \;.
\label{Hamiltonian}
\end{equation} 
The particles are assumed spinless for simplicity, and we impose periodic boundary conditions. The dispersion relation $\epsilon_{\bf k}$ is controlled by the optical lattice parameters and we will specify to a hypercubic lattice later in the calculation. The wave vectors ${\bf k}$ are contained in the first Brillouin zone.  The physical content of the repulsive mean-field interaction term $\hat{V}_{mf}=\frac{a}{2V}\hat{N}^2$ ($a>0$) is best understood by noting that it arises
from the long-range repulsive part  $v(r)$ of a 2-particle interaction potential upon performing the Kac scaling limit $\lim_{\gamma\to 0}\gamma^dv(\gamma r)$, i.e. for vanishing interaction strength and diverging range. The presence of the $1/V$ factor in $\hat{V}_{mf}$ assures extensivity of the system. The continuum version of the model in the bulk  was studied in Refs.~\onlinecite{Davies_1972, Buffet_1983, Zagrebnov_2001, Jakubczyk_2013_2}. The finite-size effects were addressed in Refs.~\onlinecite{Napiorkowski_2011, Napiorkowski_2013, Diehl_2017, Jakubczyk_2016_2}. The setup involving a harmonic trap was considered in Ref.~\onlinecite{Mysliwy_2019}. Before proceeding, we invoke results of high relevance to the present study: as was established in Ref.~\onlinecite{Napiorkowski_2013}, the Bose-Einstein condensation in the (isotropic) imperfect Bose gas is controlled by the same critical exponents as the spherical model, which in turn belongs to the bulk universality class of the $O(N\to\infty)$ model. However, the scaling function for the excess surface free energy obtained in Ref.~\onlinecite{Napiorkowski_2013} turned out to differ from its counterpart in the spherical model\cite{Dantchev_1996} by a global factor of two. This issue was further analyzed in Ref.~\onlinecite{Diehl_2017}, which established an equivalence between the isotropic imperfect Bose gas and the $O(2N)$ model for $N\to\infty$ providing a resolution of the puzzle. 
 The key features of the phase diagram and correlation functions of the imperfect Bose gas in presence of anisotropies were addressed in Ref.~\onlinecite{Jakubczyk_2018}.  The following part of the present section is a brief summary of some aspects of that study and Sec.~IV constitutes its extension accounting for the interplay between the different terms of $\epsilon_{\bf k}$ giving rise to the dimensional crossovers. 
We begin with the expression for the grand canonical partition function\cite{Napiorkowski_2011, Jakubczyk_2018} 
\begin{equation}
\label{partition}
\Xi(\mu,V,T)=-i\exp\bigg(\frac{\beta V}{2a} \mu^2\bigg)\sqrt{\frac{V}{2 \pi \beta a}}\int_{\beta \alpha - i \infty}^{\beta \alpha + i \infty} \di s \, \exp[-V \varphi(s)]\;,
\end{equation}
where $\alpha<0$ is arbitrary, $\beta^{-1}=k_B T$ and 
\begin{equation}
\varphi(s)= \frac{1}{\beta a}\bigg(-\frac{s^2}{2}+s\beta \mu\bigg)-\frac{1}{V}\log \Xi_0\bigg(\frac{s}{\beta},T\bigg)\;.
\end{equation}
The quantity $\Xi_0\big(\frac{s}{\beta},T\big)$ is the grand canonical partition function of the noninteracting Bose gas\cite{Ziff_1977} evaluated at chemical potential $\mu=\frac{s}{\beta}$ and temperature $T$. The presence of the volume factor in the term $\exp[-V \varphi(s)]$ in Eq.~(\ref{partition}) implies that the saddle point treatment of Eq.~(\ref{partition}) becomes exact in the thermodynamic limit. The saddle-point equation $\varphi'(s=s_0)=0$ yields
\begin{equation}
\label{crucial}
-s_0 \frac{1}{a \beta} +\frac{\mu}{a}= \frac{1}{V}\sum_{n=1}^{\infty}e^{ns_0}\sum_{\textbf{k} \neq 0} e^{-n \beta \epsilon_{\textbf{k}}} +\frac{1}{V} \frac{e^{s_0}}{1-e^{s_0}}\;.
\end{equation} 
It is crucial for exploiting the thermodynamics of the system and the whole analysis to follow (see Ref.~\onlinecite{Napiorkowski_2011} for detailed explanations in the isotropic case). 
As demonstrated in Ref.~\onlinecite{Jakubczyk_2018}, the thermodynamic properties of the system in the vicinity of the critical temperature (and for $T$ low enough) are fully determined by the asymptotic form of the dispersion relation $\epsilon_{\textbf{k}}$ at $|\textbf{k}|$ small. The system displays a line of second order phase transitions $T_c(\mu)$ down to $T_c(\mu\to 0)\to 0$. 

Considering the hypercubic lattice as a specific example, we  take
 \begin{equation}
\label{anis}
\eps_{\textbf{k}}=\sum_{\vR}2 \,t_{\vR} \,\big(1-\cos(\vk \cdot \vR)\big)\;,
\end{equation}
where ${\bf R}$ runs through the Bravais lattice.
 In a typical situation, expansion around ${\textbf{k}}=0$ leads to the following asymptotic form 
 \beq
 \label{tildeeps}
 \eps_{\textbf{k}} \rightarrow \tilde{\eps_{\textbf{k}} }=\sum_{i=1}^{d} c_i (k_i A) ^2\;,
 \eeq
 where $c_i$ are numerical coefficients, and $A$ denotes the lattice constant. From the point of view of universal properties, the behavior of the system characterized by the dispersion of  Eq.~(\ref{tildeeps}) is identical to that of the continuum imperfect Bose gas. By tuning the hopping parameters it is however possible to cancel one or more of the coefficients $c_i$ in which case, the corresponding leading order term in the $i$-th direction becomes quartic (or even higher order). Specific examples of such a tuning procedure are given in Ref.~\onlinecite{Jakubczyk_2018}. In a general situation, the asymptotic form of the dispersion may be written in the form given by Eq.~(\ref{tildeeps2}).
 The thermodynamics and correlations of the system characterized by the dispersion of Eq.~(\ref{tildeeps2}) were thoroughly studied in Ref.~\onlinecite{Jakubczyk_2018} which pointed at the affinity to the isotropic system in an effective dimensionality 
 \begin{equation} 
 d_{eff}=\frac{2}{\psi}\;, 
 \end{equation} 
 where 
 \beq
 \label{psi}
\frac{1}{\psi}=\frac{1}{\alpha_1}+ \ldots + \frac{1}{\alpha_d}\;.
 \eeq
 In particular, the system is above its lower critical dimension (and therefore hosts a Bose-Einstein condensed phase in its phase diagram at $T>0$) if $\frac{1}{\psi} > 1$. The upper critical dimension is on the other hand determined by the condition $\frac{1}{\psi}=2$. We also remark\cite{Jakubczyk_2018} that the asymptotic shape of the critical line in the phase diagram is given by the universal exponent $\frac{1}{\psi}$ so that $\mu_c (T) \sim T^{\frac{1}{\psi}}$ (for the continuum case one has $\mu_c (T) \sim T^{d/2}$). We point out that overall the role played in the continuum case by the spatial dimensionality $d$ is for the optical lattice taken over by the parameter $\frac{2}{\psi}$. This may in turn be experimentally tuned leading to crossovers between different effective dimensionalities involving also fractional values. The analysis of such crossover effects requires, however, going beyond the asymptotic form of $\eps_{\textbf{k}}$ given by Eq.~(\ref{tildeeps2}) and accounting for the next-to leading contributions. We present such an extension below.   
 \section{Dimensional crossover} 
 In the most standard setup, the dimensional crossover is realized by confining the system in one or more directions and manipulating the thermodynamic parameters so that the characteristic length scale becomes larger (or smaller) than the confining parameter [see e.g. \onlinecite{Lammers_2016, Delfino_2017, Ilg_2018, Faigle_2019, Noh_2020}],  Here we analyze a natural alternative avenue, already outlined in Sec.~III, where the dimensional crossover between distinct effective dimensionalities is tuned by manipulating the hopping parameters.  
 To this aim we now extend the expansion of the dispersion given by Eq.~(\ref{tildeeps2}) and consider 
 \begin{equation}
\label{siec1}
\tilde{\eps_{\textbf{k}} }=t_0 (k_1A)^2+t(k_2 A)^{2m} +\tau (k_3 A)^2+ \tau ' (k_3 A)^4\;, 
\end{equation}
where $t_0,\,t,\, \tau, \,\tau'>0$. This allows us to analyze  different physical situations depending, in particular, on the value of $m$. On the other hand, the fixed signs of the kinetic couplings restrict to uniform ordered phases, ruling out the modulated states related to the Lifshitz points.\cite{Diehl_2002} For the time being we specified  to $d=3$. We will restore the generality of $d$ in the discussion of the Casimir effect (Sec.~V). 

With $\tilde{\eps_{\textbf{k}} }$ given above, the saddle-point equation may be written as  
\begin{align}
-s_0 \frac{1}{a \beta} + \frac{\mu}{a}&=  \frac{1}{A^3} \frac{\Gamma(1+\frac{1}{2m})}{ 2^{3/2} \, \pi ^{5/2} } \frac{1}{ \beta ^{1+\frac{1}{2m}}  \, \sqrt{t_0 \, \tau  \, t^{1/m}}} \sum_{n=1}^{\infty} \frac{e^{n s_0}}{n^{1+\frac{1}{2m}}} f(n \theta) \nonumber \\ &+\frac{1}{V} \frac{e^{s_0}}{1-e^{s_0}}\;, 
\label{speqq}
\end{align}
where:
\begin{equation}
f(x)=\sqrt{x}  e^x  K_{1/4}(x)\;, \;\;\; \theta=\frac{\beta \tau ^2}{8 \tau'}\;,
\end{equation}
and $K_{\alpha}(x)$ is the Bessel function.
As we show below, the dimensionless parameter $\theta$ serves as the scaling variable controlling the dimensional crossover. Note that it may be varied between 0 and infinity either by manipulating the hopping parameters, or temperature. The function $f(x)$ is monotonously increasing and bounded. Its asymptotic behavior is given by  
\begin{equation}
\label{f0}
f(x) \xrightarrow{x \to 0^+} \frac{\Gamma(1/4) x^{1/4}}{2^{3/4}}\;,\;\;\;\; f(x) \xrightarrow{x \to \infty} \sqrt{\frac{\pi}{2}}\;.
\end{equation} 
The expression for the critical line is obtained\cite{Napiorkowski_2011} by dropping the last term in Eq.~(\ref{speqq}) and putting $s_0=0$. It reads: 
\begin{equation}
\label{crit}
\mu_c (T) = \frac{a}{A^3} \frac{\Gamma(1+\frac{1}{2m})}{2^{3/2} \, \pi ^{5/2}} \frac{1}{ \beta ^{1+\frac{1}{2m}}  \, \sqrt{t_0 \, \tau  \, t^{1/m}}} \sum_{n=1}^{\infty} \frac{1}{n^{1+\frac{1}{2m}}} f(n \theta)\;.
\end{equation} 
The phase hosting the condensate is stable for $\mu>\mu_c(T)$. 
We now analyze the properties of $\mu_c (T)$ focusing on two very different cases.  In the first considered example the crossover occurs between two effective dimensionalities both located above the lower critical dimension $d_l$. In the second case, the system is tuned to the effective dimensionality exactly at $d_l$, and the condensate is marginally depleted from the system. We subsequently discuss the general situation. 
 \subsection{Case $\tilde{\eps_{\textbf{k}} }=t_0 (k_1A)^2+t_0(k_2 A)^2 +\tau (k_3 A)^2+ \tau ' (k_3 A)^4$}
 Here we consider $m=1$ and analyze the crossover in the effective dimensionality realized by changing $\theta$. For example we may vary the parameter $\tau>0$ towards zero, gradually giving way to the subdominant term proportional to $k_3^4$ in the dispersion along the 3rd direction. The relevant values of $1/\psi$ are $1/\psi = 3/2$ (for $\tau>0$) and $1/\psi=5/4$ (for $\tau=0$). 
 The series in Eq.~(\ref{crit}) is convergent for any $\theta>0$. At fixed $\tau$ and $\tau'$ the variable $\theta$ may be tuned between the asymptotic regimes $\theta\gg 1$ and $\theta\ll 1$ by varying temperature $T$. Alternatively, at given $T$ (and $\tau'$), one may use $\tau$ as the control parameter. In the asymptotic regime $\theta\gg 1$ we may replace $f(n\theta)$ by its limiting form for large arguments. This leads to 
\begin{equation}
\label{siec1l}
\mu_c^{\infty}(T)  \approx \frac{1}{8 \pi ^{3/2}}\, \zeta\bigg(\frac{3}{2}\bigg) \, \frac{a}{A^3} \, \frac{1}{ t_0 \sqrt{\tau}} \, (k_B T)^{3/2}\;,
\end{equation}
where $\zeta \left(z\right)=\sum_{n=1}^{\infty}\frac{1}{n^{z}}$ (with $z>1$) is the Riemann zeta function. The expression of Eq.~(\ref{siec1l}) coincides with the formula for $T_c$ derived in Ref.~\onlinecite{Jakubczyk_2018}. 

In the opposite limit $\theta\ll 1$ the series is dominated by terms with $n\theta\ll 1 $ and one may use asymptotic form of the function $f$ for small arguments. This leads to the expression 
 \begin{equation}
\label{siec1h}
\mu_c^{0}(T)\approx\frac{\Gamma(1/4)}{16 \pi ^2} \zeta \bigg(\frac{5}{4}\bigg) \frac{a}{A^3} \frac{1}{t_0 \sqrt[4]{\tau'}} (k_BT)^{5/4}\;, 
\end{equation} 
which also agrees with the predictions of Ref.~\onlinecite{Jakubczyk_2018}.  
Estimating the crossover temperature $T_{cross}$ by the condition $\mu_c^{\infty}(T_{cross})=\mu_c^{0}(T_{cross})$, we find 
\begin{equation}
\label{cross1}
k_B T_{cross}=\frac{\tau ^2}{\tau '}\Bigg( \frac{\Gamma(1/4)}{2 \sqrt{\pi}} \frac{\zeta(\frac{5}{4})}{\zeta(\frac{3}{2})}\Bigg)^4.
\end{equation} 
The emergent picture is illustrated in Fig.~1. 
\begin{figure}[ht]
\begin{center}
\label{}
\includegraphics[width=7.5cm]{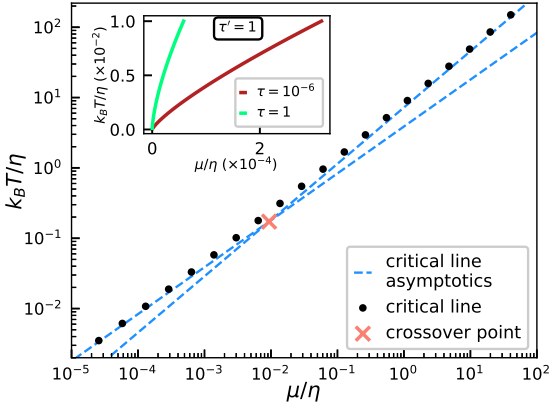}
\caption{The critical line as computed from Eq.~(\ref{crit}) with $m=1$. The plot parameters are $t_0=t=1$, $\tau=0.2$, $\tau'=2.43$. The plotted quantities are scaled by $\eta=\frac{a}{A^3}$. One finds the crossover between the scaling behavior with $\psi=2/3$ (low $T$) and $\psi=4/5$ (high $T$). The crossover temperature precisely agrees with Eq.~(\ref{cross1}) (red cross). The inset presents the critical line in linear scale for $\tau=0$ (and therefore $\theta=0$) and $\tau=1$, where the scaling with only one value of $\psi$ (4/5 or 2/3) occurs in the entire temperature range. }
\end{center}
\end{figure} 
While the asymptotic scaling behavior is clear from the earlier studies of Ref.~\onlinecite{Jakubczyk_2018}, the identification of the scaling variable $\theta$, and, in consequence, also the crossover scale is possible only by considering the extended dispersion of Eq.~(\ref{siec1}). The entire picture becomes less obvious also from the point of view of the asymptotic scaling of $T_c$ when one of the effective dimensionalities is at the lower critical dimension $d_l$ for condensation. We analyze this situation below in Sec.~IVB.

 \subsection{Case $\tilde{\eps_{\textbf{k}} }=t_0(k_1A)^2+t(k_2A)^4+\tau (k_3 A)^2+ \tau ' (k_3 A)^4$}
 The present case of $m=2$ in Eq.~(\ref{crit}) is special in that the value of $1/\psi$ obtained for $\tau\to 0$ corresponds to the effective lower critical dimension $d_{eff}=d_l=2$.  The condensate is therefore marginally unstable for $\tau\to 0$ at finite temperatures. We consider the setup where $\tau$ is gradually switched off and follow the way the condensate becomes expelled from the phase diagram. The kinetic term is therefore used as the control parameter tuning the system across the phase transition.  
 
 The series in Eq.~(\ref{crit}) is again convergent for any $\theta>0$. For $\theta\gg 1$ we recover the expression given by Eq.~(\ref{siec1h}) along the line analogous to Sec.~IVA.  However, in contrast to the case considered in Sec.~IVA if the function $f(x)$ is replaced by its asymptotic form for small arguments, one obtains a divergent expression (which is a reflection of the absence of condensation for $\tau=0$). 
 
 We now describe a procedure to  extract the asymptotic behavior of the critical line at $\theta\ll 1$. For this purpose we use Eq.~(\ref{f0}) and split the series in Eq.~(\ref{crit}) as follows:   
 \begin{equation}
\label{series}
\sum _{n=1}^{\infty}\frac{1}{n^{5/4}} \; f(n\theta) \approx \sum_{n=1}^{N(\theta)}\frac{1}{n}\frac{\Gamma(1/4)\theta ^{1/4}}{2^{3/4}}+\sum_{N(\theta)+1}^{\infty}\frac{1}{n^{5/4}}\; f(n\theta)\;.
\end{equation}
The auxiliary parameter $N(\theta)\approx \tilde{\alpha}/\theta$ (with a numerical constant $\tilde{\alpha}$) should correspond to a value such that (at $\theta>0$ fixed) the first term in Eq.~(\ref{series}) constitutes a valid approximation to the entire series. The last term of Eq.~(\ref{series}) is bounded from above by:
\begin{equation}
\sum_{N(\theta)+1}^{\infty}\frac{1}{n^{5/4}}\; f(n\theta) < \sqrt{\frac{\pi}{2}} \frac{4}{\sqrt[4]{N(\theta)}}\;.
\end{equation}
 The upper bound obviously vanishes for $N(\theta)\to\infty$. The first term on the right-hand side (RHS) of Eq.(\ref{series}) may be estimated using the Euler formula
 \begin{equation}
\sum_{n=1}^{N(\theta)}\frac{1}{n} \approx \log N(\theta) +\gamma +\frac{1}{2 N(\theta)}+O\bigg(\frac{1}{N(\theta)^2}\bigg)
\end{equation}
with $\gamma$ denoting the Euler-Mascheroni constant. Truncating this expansion at the leading term and using Eq.~(\ref{series}), the asymptotic form of Eq.~(\ref{crit}) becomes
 \begin{equation}
\label{lat2zero}
\mu_c^{0}(T) \approx \frac{\Gamma(1/4)^2}{32 \pi^{5/2}} \frac{a}{A^3} \frac{1}{\sqrt{t_0} \; t^{1/4} \sqrt[4]{\tau'}}\, \log \bigg( \frac{1}{\theta}\bigg) (k_B T)\;.
\end{equation}
 Note that the unspecified constant $\tilde{\alpha}$ [relating $\theta$ and $N(\theta)$] as well as the constant $\gamma$ influence only  the subdominant contribution to $\mu_c^{0}(T)$, which is also linear in $T$, but does not involve the log-divergent coefficient $\sim \log ( \frac{1}{\theta})$. 
 The above calculation reveals a somewhat subtle behavior of the critical line $T_c(\mu)$. As we have shown, at \emph{fixed} $\theta$ the dependence $T_c(\mu)$ is linear for $\mu$ large  and, at $\mu$ smaller, it crosses over to the power law behavior with the exponent $\psi=4/5$. When the parameter $\tau$ is then tuned towards zero (implying vanishing $\theta$), the coefficient governing the high-$\mu$ (high $T_c$) linear behavior vanishes logarithmically, thus suppressing the critical temperature towards zero. This is accompanied by shifting the scale corresponding to the onset of the power-law regime towards zero chemical potentials. This picture clarifies the mechanism leading to continuously depleting the condensate from the system for $\tau\to 0$. The corresponding illustration is presented in Fig.~2
 \begin{figure}[ht]
\begin{center}
\label{}
\includegraphics[width=7.5cm]{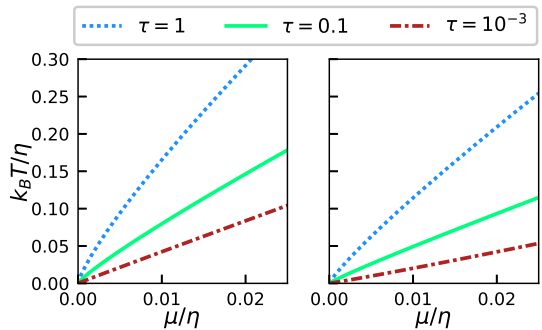}
\caption{The critical line as computed from Eq.~(\ref{crit}) with $m=2$ (left panel) and $m=3$ (right panel). The plot parameters are $t_0=t=\tau'=1$ and the plotted quantities are scaled by $\eta=\frac{a}{A^3}$. In the former case one finds the crossover between the scaling behavior with $\psi=4/5$ (low $T$) and the essentially linear behavior (high $T$). Upon tuning $\tau$ towards zero, the coefficient governing the high-$T$ behavior vanishes logarithmically and the scale of $\mu$ corresponding to the onset of the behavior with $\psi=4/5$ shifts towards zero. In the latter case ($m=3$) the linear coefficient of the $T_c$-line vanishes algebraically (see the main text).   }
\end{center}
\end{figure} 

Concerning the situation with a value of $m$, where the effective dimensionality $d_{eff}$ is below $d_l$ for $\tau=0$, one may show that the picture is similar to the one extracted above for $m=2$.  The critical line is linear in the high-$T$ regime [$T_c(\mu)\approx \tilde{A}\mu$] and crosses over to a power-law with an exponent $\psi<1$ at $T$ smaller. The role of $m$ reveals itself in the way the coefficient $\tilde{A}$ governing this linear behavior vanishes for $\tau\to 0$. Instead of the behavior $\tilde{A}\sim -1/\log(\tau)$ obtained above for $m=2$, one finds a power-law dependence $\tilde{A}\sim\tau^{\kappa(m)}$. For $m=3$ we obtain $\kappa(3)=1/6$. 
 
We finally point out an observation concerning the relation to a general situation with short-ranged interaction potentials and suggesting that the obtained picture may be valid also for non-mean-field models. As we already remarked, the bulk critical exponents controlling Bose-Einstein condensation in the imperfect Bose gas are known to be the same as the Berlin-Kac (spherical) model,\cite{Berlin_1952} which in turn corresponds to the limit  $N\to\infty$ of the $O(N)$-symmetric models.\cite{Stanley_1968} The renormalization-group studies of the quantum variants of the $O(N)$ models (see e.g. \onlinecite{Millis_1993, Bijlsma_1996, Andersen_1999, Crisan_2002, Nikolic_2007, Jakubczyk_2008, Jakubczyk_2009_phi6, Jakubczyk_2010}) reveal that the universal aspects of the $T_c$-line (the shift  exponent $\psi$ in particular) are insensitive to the value of $N$, but instead are fully determined by the spatial dimensionality $d$ and the dynamical exponent $z$ via the simple relation 
\begin{equation}
\psi=\frac{z}{d+z-2}  \;.
\end{equation} 
For interacting bosons one has $z=2$ and therefore $\psi=2/d$ which fully agrees with the expression obtained for the imperfect Bose gas as a result of an exact analysis. The equivalence holds also for the upper and lower critical dimensions (also for cases with modified dispersions). It is natural to conjecture that as far as the universal properties are concerned, the entire picture derived in this section remains unchanged if one replaces the Kac-scaled interaction potential with a (more realistic) short-ranged interaction (or, in other words, the entire picture retains the independence on the value on $N$ also for the essential features of the crossover behavior). A verification of this hypothesis requires further studies from the renormalization-group point of view, which we leave to future work.
 
 \section{Casimir energy}
We now move on to discuss the thermodynamic Casimir effect in the system. We consider a general situation where the $d$-dimensional system is enclosed in volume $V=L^{d-1}D$, where $L\gg D\gg l_{mic}$ and $l_{mic}$ denotes all the microscopic length scales present in the system. The quantity $D$ measures the system extension in the $d$-th direction (which is a 'special' direction - compare Sec.~II). We will separately consider the situation with the confining walls perpendicular to ${\bf k}_1$ ('normal' direction) in Sec.~VD.  We  analyze the case of periodic boundary conditions.
The dispersion displays $\sim k^4$ behavior in $m<d$ directions, and the usual $\sim k^2$ behavior in the remaining $d-m$ directions and its asymptotics is given by Eq.~(\ref{asssym}).
 Importantly, the dispersion parameters ($t_0$ and $t$) are assumed to be the same (i.e. independent of $i$) for each of the two classes of spatial directions. Relaxing this symmetry may lead to additional effects\cite{Diehl_2003} not addressed here. 
We keep only the dominant contributions in each of the directions, leaving the crossover effects aside. We also introduce $\tilde{\epsilon}_{k_1}=t_0 (k_1 A)^2$ and $\tilde{\epsilon}_{k_d}=t (k_d A)^4$.  We are interested in the excess grand-canonical free energy density 
\begin{equation}
\label{excess}
\omega_s(D,\mu,T)= \lim_{L\to\infty}\left[\frac{\Omega(L,D,T,\mu)}{L^{d-1}}-D\omega_b(T,\mu)\right]
\end{equation}
which is related to the Casimir force $F(D,\mu,T)$ by $F(D,\mu,T)=-\frac{\partial \omega_s(D,\mu,T)}{\partial D}$. The grand-canonical free energy is given by $\Omega(L,D,T,\mu)=-\beta^{-1}\ln \Xi(T,L,D,\mu)$ and the bulk free energy density $\omega_b(T,\mu)$ follows from $\omega_b(T,\mu)=\lim_{L\to\infty}\frac{1}{L^d}\Omega(L,D=L,T,\mu)$. Using Eq. \eqref{partition}, the excess contribution to the grand potential can be written as 
\begin{equation} 
\label{omegas}
\omega_s(D,\mu,T)=\lim_{L\to\infty}\beta^{-1}D\left[\varphi (\bar{s})-\varphi_b(s_0)\right]\;,
\end{equation}
where 
\begin{align} 
\label{phi_def}
\varphi (\bar{s})=&-\frac{\bar{s}^2}{2 a\beta}+\frac{\mu \bar{s}}{a}- \nonumber\\
&\frac{1}{V}\left[\sum_{\textbf{k}\neq({\bf 0}, k_d)}\sum_{r=1}^{\infty}\frac{1}{r}e^{r(\bar{s}-\beta\tilde{\eps_{\textbf{k}}})}-\sum_{k_d}\log\left(1-e^{\bar{s}-\beta\tilde{\epsilon}_{k_d}}\right)\right]\;,
\end{align}
$\bar{s}$ represents the solution to the saddle-point equation $\varphi'(\bar{s})=0$, $s_0$ corresponds to $\bar{s}$ in the bulk case (i.e. when $D=L$ and $L\to\infty$) and $\varphi_b(s)=\lim_{D\to L}\varphi (s)$. In essence, our present goal amounts to solving the saddle-point equation at finite $D$ and evaluating Eq.~(\ref{omegas}). 

We identify two distinct thermal length scales  
\begin{equation}
\lambda_1 = 2A \, \sqrt{ \pi} \, \sqrt{\beta \, t_0} \qquad \lambda_2=A \, \frac{\pi}{\Gamma(5/4)} \, (\beta \, t)^{1/4}\;,
\end{equation}
conveniently absorbing numerical factors. The scales $\lambda_1$ and $\lambda_2$ are analogous to the thermal de Broglie length of the isotropic continuum gas and correspond to normal and special directions, respectively. We assume them to be large as compared to the lattice scale $A$. Note that $\lambda_1$, $\lambda_2$ diverge for $T\to0$. We will later assume that $D/\lambda_1\gg 1$, $D/\lambda_2\gg 1$. This excludes considering the limit $T\to 0$, which is beyond the scope of the present paper. 
\subsection{Bulk limit }
The bulk saddle-point equation (where $D=L\to\infty$) can be written as 
\begin{equation}
\label{seqcasb}
-s_0 \frac{1}{a \beta} +\frac{\mu}{a}=\frac{1}{\lambda_1^{d-m} \, \lambda_2^{m}}g_{\frac{1}{\psi}}(e^{s_0})+ \frac{1}{V} \frac{e^{s_0}}{1-e^{s_0}}
\end{equation}
with 
\begin{equation}
\frac{1}{\psi}=\frac{d}{2}-\frac{m}{4}
\end{equation}
and the Bose function $g_n(z)=\sum_{k=1}^\infty \frac{z^k}{k^n}$. We read off the expression for the critical line
\begin{equation}
\mu_c(T)=\frac{a}{\lambda_1^{d-m} \, \lambda_2^{m}}\zeta\Big( \frac{1}{\psi} \Big)
\end{equation}
recovering the previously studied behavior $\mu_c(T) \sim T^{1/\psi}$. Introducing the dimensionless parameter measuring the distance from the  bulk critical line
\begin{equation}
\varepsilon=\frac{\mu -\mu_c}{\mu_c}\
\end{equation}
and  expanding the Bose function for $|s_0|\ll1$ according to 
\begin{equation}
    g_{\epsi}(e^{s_0})-\zeta \Big(\epsi \Big) \approx \begin{cases}
        \Gamma \big(1-\epsi\big)|s_0|^{\epsi-1}, &  1<\epsi<2 \\
        |s_0|\log|s_0|, &  \epsi=2\\
        -\zeta\big(\epsi-1\big)|s_0|, &  \epsi>2
        \end{cases} 
\label{Bose_asymptotics}        
\end{equation}
we may solve Eq.~(\ref{seqcasb}) for $|\varepsilon|\ll1$ and obtain $\varphi_b(s_0)$. The analysis of the bulk limit then proceeds along the line of Refs.~\onlinecite{Napiorkowski_2013, Jakubczyk_2018}.
 \subsection{Saddle-point equation}
 We now analyze the situation, where the system remains finite in one of the directions so that $L\to \infty$, but $D$ (i.e. the system extension in the $d$-th direction) is kept finite. The saddle-point equation is first cast in the form 
\begin{equation} 
\begin{split}
\zeta \Big(\epsi\Big) \bigg(-\frac{\bar{s}}{\mu_c \beta}+\varepsilon\bigg)= &-\zeta \Big(\epsi\Big)+\frac{\lambda_2}{D}\sum_{r=1}^{\infty}\frac{e^{r \bar{s}} }{r^{\epsi-\frac{1}{4} }}\sum _{k_d}e^{-r \beta \tilde{\epsilon}_{k_d} }\\ 
&-\frac{\lambda_1^{d-m} \lambda_2^{m}}{V} \sum_{k_d} \frac{1 }{1-e^{\beta \tilde{\epsilon}_{k_d} -\bar{s}}}\;.
\end{split} 
\label{speq32}
\end{equation}
The sum occurring in the second term on the RHS of the equation can be transformed using the Poisson formula 
\beq 
\sum_{m=-\infty}^{\infty} f(m)=\sum_{n=-\infty}^{\infty} \hat{f}(n)\;, 
\eeq
where  $\hat{f}(n)=\int_{-\infty}^{\infty} \di x \, e^{-i2 \pi n x}  \,f(x)$. We obtain 
\begin{equation} 
\label{k_d_sum}
\sum_{k_d} e^{-r\beta \tilde{\epsilon}_{k_d}}=\frac{D}{2\Gamma(5/4)\lambda_2 r^{1/4}} \sum_{n=-\infty}^{\infty} \phi \Big( \frac{\pi n}{\Gamma(5/4)} \frac{D}{\lambda_2} \frac{1}{r^{1/4}} \Big)\;,
\end{equation}
where 
\begin{equation}
\label{phi_function}
\phi(k)= \int_{-\infty}^{\infty} \di x \, e^{i k x} \, e^{-x^4}\;.
\end{equation}
 The properties of the function $\phi(k)$ are crucial for the results to follow. In particular $\lim_{k \to 0} \phi(k)=2 \Gamma(5/4)$, while the asymptotic behavior at $k$ large is described by\cite{Boyd_2014} 
\begin{equation}
\phi(k) \sim 2^{7/6}\sqrt{\frac{\pi}{3}} \frac{1}{k^{4/3}} \exp \left(-\frac{3}{16} 2^{1/3} k^{4/3}\right) \cos \left(\frac{3^{3/2} 2^{1/3} }{16} k^{4/3} -\frac{\pi}{6}\right)
\end{equation}
so that it exhibits exponentially damped oscillations.  This gives rise to substantial differences as compared to the usual case with quadratic $\tilde{\epsilon}_{k_d}$, where the corresponding expression is a monotonously decreasing Gaussian function. An illustrative plot of $\phi(k)$ is given in Fig.~3 (left panel). 
\begin{figure}[ht]
\begin{center}
\label{}
\includegraphics[width=7.5cm]{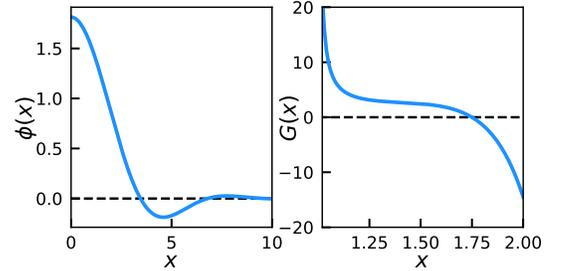}
\caption{The functions $\phi(x)$ and $G(x)$, see the main text.    }
\end{center}
\end{figure} 
 We now consider $D\gg\lambda_2$ and replace the summation over $r$ in Eq.~(\ref{speq32}) with an integral in accord with the Euler-Maclaurin formula. 
We introduce the following notation
\begin{equation}
\sigma=\frac{\pi}{\Gamma(5/4)} \frac{D}{\lambda_2} |\bar{s}|^{1/4} \qquad F_{\kappa} (x) = \int_0^{\infty} \di p \, \frac{e^{-p}}{p^\kappa} \phi (x/p^{1/4})
\end{equation}
and transform the saddle-point equation to the following form 
\begin{equation}
\begin{aligned}
&\zeta \Big(\epsi\Big) \bigg(-\frac{\bar{s}}{\mu_c \beta}+\varepsilon\bigg)=g_{\epsi}(e^{\bar{s}})-\zeta \Big(\epsi\Big)+\\ 
&\frac{\Gamma(5/4)^{\frac{4}{\psi}-5}}{\pi^{\frac{4}{\psi} -4}} \Big ( \frac{\lambda_2}{D} \Big)^{\frac{4}{\psi}-4} \sigma^{\frac{4}{\psi}-4} \sum _{n=1}^{\infty} F_{\epsi}(n \sigma) 
-\frac{\lambda_1^{d-m} \lambda_2^{m}}{V} \sum_{\vkd} \frac{1 }{1-e^{\beta \tilde{\epsilon}_{\vkd} -\bar{s}}}. 
\end{aligned}
\label{speq}
\end{equation} 
The function $F_{\kappa} (x)$ is characterized by oscillatory behavior inherited from $\phi(k)$.
In the subsequent step we expand the Bose function for small $\bar{s}$ according to Eq.~(\ref{Bose_asymptotics}) and perform the limit $L\to\infty$. We also introduce the scaling variable $x$
\begin{equation}
x=\begin{cases}
        \varepsilon \Big(\frac{D}{\lambda_2} \Big)^{\frac{4}{\psi}-4}, &  1<\epsi<2 \\
        \varepsilon \Big(\frac{D}{\lambda_2} \Big)^4, &  \epsi >2\;,
        \end{cases}
\end{equation} 
which is positive below bulk $T_c$ and negative otherwise.
This allows us to write Eq.~(\ref{speq}) as a transparent relation between the variable $x$ and $\sigma=\sigma(x)$. As will turn out, the dependence of the scaling function for the Casimir energy on $\bar{s}$ can be absorbed into $\sigma$. For $1<\epsi<2$ we obtain 
\begin{align}
\label{eqsadle}
\zeta \Big(\epsi\Big) x = &\bigg( \frac{\Gamma(5/4)}{\pi} \bigg)^{\frac{4}{\psi}-4} \sigma^{\frac{4}{\psi}-4}\bigg[ \Gamma \Big(1-\epsi \Big)+\frac{1}{\Gamma(5/4)} \sum_{n=1}^{\infty} F_{\epsi} (n \sigma) \bigg]  \nonumber   \\
+      &  \frac{\lambda_1^{d-m} \lambda_2^{m}}{V} \left(\frac{D}{\lambda_2}\right)^{\frac{4}{\psi}-4}  \sum_{\vkd} \frac{1 }{e^{\beta \tilde{\epsilon}_{\vkd} -\bar{s}}-1}\;,
\end{align}
while for $\epsi >2$ we find 
\begin{equation}
\label{sad1}
\begin{aligned}
&\zeta \Big(\epsi \Big) x = - \bigg( \frac{\Gamma(5/4)}{\pi} \bigg)^4 \sigma^4 \Bigg[ \frac{ \zeta(\epsi) }{\mu_c \beta} +\zeta \Big( \epsi-1\Big) \Bigg]+\\
&+ \frac{1}{\Gamma(5/4)} \bigg( \frac{\Gamma(5/4)}{\pi} \bigg)^{\frac{4}{\psi}-4} \Big( \frac{\lambda_2}{D} \Big)^{\frac{4}{\psi}-8}\sigma^{\frac{4}{\psi}-4} \sum_{n=1}^{\infty} F_{\epsi}(n \sigma) + \\
& \frac{\lambda_1^{d-m} \lambda_2^{m}}{V} \left(\frac{D}{\lambda_2}\right)^4 \sum_{\vkd} \frac{1 }{e^{\beta \tilde{\epsilon}_{\vkd} -\bar{s}}-1} \;.
\end{aligned}
\end{equation}
In each of the cases, the dependence on $x$ occurs only on the left-hand side (LHS) of the equation, while the dependence on $\sigma$ on the corresponding RHS. Additionally, let us notice that the last term with $\textbf{k}_d=0$ resembles the last term of Eq. \eqref{crucial}, which in turn, in the bulk limit, is proportional to the condensate density~\cite{Napiorkowski_2011}. We discuss the saddle-point solution in the two cases separately. We leave aside the case $\frac{1}{\psi}=2$, corresponding to the upper critical dimension where logarithmic corrections arise [see Eq.~(\ref{Bose_asymptotics})], but apart from them the behavior of the scaling function is expected to be very similar. 
\subsubsection{Case $1<\frac{1}{\psi}<2$}
Let us first concentrate on the low-$T$ phase ($x \geq 0$). One may show that the RHS of Eq.~(\ref{eqsadle}) as a function of $\sigma$ is unbounded from above for $\frac{1}{\psi}\leq \frac{5}{4}$. This follows from the properties of the function $F_{\kappa}(x)$.  In such a situation for each $x\geq0$ one finds a unique $\sigma(x)>0$. In consequence, the corresponding value of $|\bar{s}|$ is controlled by $D$ (i. e. vanishes for $D\to \infty$)  and the last term of Eq.~(\ref{eqsadle}) vanishes for $L\to\infty$. The situation is more complex for   $\frac{1}{\psi}> \frac{5}{4}$. In this case the $L$-independent term on the RHS of Eq.~(\ref{eqsadle}) is bounded from above by its value at $\sigma\to 0^+$, which in turn may be expressed as  $\frac{4}{\Gamma(5/4)} \left( \frac{\Gamma(5/4)}{\pi} \right)^{\frac{4}{\psi}-4} G\left(\epsi\right) \zeta\left(\frac{4}{\psi}-4\right)$, where we introduce 
\begin{equation}
G(\kappa)=\int_0^{\infty} \di q \, q^{4\kappa-5} \,\phi(q)\;.
\end{equation}
The properties of the above function are important for the analysis to follow. It is plotted in Fig.~3 (right panel) for illustration. We note in particular that
\begin{equation}
F_\kappa (x) \approx \frac{4}{x^{4\kappa-4}}G(\kappa)
\end{equation}
for $x\ll 1$ and $\kappa>1$.
The physical significance of the value $\frac{1}{\psi}=\frac{5}{4}$ is clear upon noticing that it corresponds to the lower critical dimension for condensation in a system with finite $D$ (i.e. after ''excluding'' the $d$-th direction in which the system is finite). One then finds a finite solution $\sigma(x)>0$ for $x$ fulfilling the condition
\begin{equation}
0\leq x \leq x_{cr}(\frac{1}{\psi})=\frac{1}{\zeta(\epsi)} \frac{4}{\Gamma(5/4)} \bigg( \frac{\Gamma(5/4)}{\pi} \bigg)^{\frac{4}{\psi}-4} G\Big(\epsi\Big) \zeta\Big(\frac{4}{\psi}-4\Big)\;.
\end{equation}
In the opposite situation (for $x>x_{cr}$) the last term in Eq.~(\ref{eqsadle}) gives a finite contribution in the thermodynamic limit. This reflects the phase transition taking place (at finite $D$) for
\begin{equation}
\bar{\mu}_c(T)=\mu_c(T)\left[x_{cr}(\frac{1}{\psi})\left(\frac{\lambda_2}{D}\right)^{\frac{4}{\psi}-4}+1\right]\;.
\end{equation}

 We then obtain  $\sigma (x)=0$ for $x>x_{cr}$. Inspection of the function $G\left(\epsi\right)$ - see Fig.~3 reveals however that  $G\left(\epsi\right)$ has a zero at  $\epsi=\frac{7}{4}$. For $\epsi>\frac{7}{4}$ we obtain $\sigma(x)=0$ for all $x\geq0$ in the limit $L\to\infty$. This behavior persists for $\frac{1}{\psi}>2$, as discussed in the next subsection. Note however that in the ''uniaxial'' case $m=1$ the value $\frac{1}{\psi}=\frac{7}{4}$ corresponds to the physical dimensionality $d=4$, while for $m=2$, $\frac{1}{\psi}=\frac{7}{4}$ implies and even higher value $d=\frac{9}{2}$. Obviously an experimentally meaningful value of $\frac{1}{\psi}$ is $\frac{5}{4}$.

 As explained above, for $x\geq0$ the behavior of $\sigma $ is controlled by either $D$ or the system volume. The situation is different for $x<0$, where it is governed by the distance from the phase transition. Indeed, fixing $x<0$ and passing to the limit $L\to\infty$, $D\to\infty$ we obtain a finite solution for $\sigma (x)$, which, for large $|x|$ (where we may replace $\bar{s}$ with bulk saddle-point $s_0<0$) is given by the relation
\begin{equation}
\zeta \left(\epsi\right) x =  \left(\frac{\Gamma(5/4)}{\pi} \right)^{\frac{4}{\psi}-4} \sigma^{\frac{4}{\psi}-4} \Gamma \left(1-\epsi \right).
\end{equation} 
\subsubsection{Case $\frac{1}{\psi}>2$}
For $\frac{1}{\psi}>2$ the saddle-point equation [Eq.~(\ref{sad1})] has a finite solution 
\begin{equation}
\sigma (x)= \frac{\pi}{\Gamma(5/4)}\Bigg( \frac{|x|}{\frac{1}{\mu_c \beta} +\zeta (\epsi-1)/\zeta(\epsi)} \Bigg)^{1/4}
\end{equation}
for $x\leq0$ in the limit $L\to\infty$, $D\to\infty$. For $x>0$ inspection of the signs of the different terms in Eq.~(\ref{sad1}) leads directly to the conclusion that this equation is never fulfilled for $\sigma (x)>0$. In consequence, the last term in Eq.~(\ref{sad1}) must give a finite contribution, which implies $\sigma (x)=0$ for $x>0$ and $L\to\infty$.

\subsection{Excess free energy}
We proceed to determine the excess grand canonical free energy given by Eq.~(\ref{excess}). We again analyze the two cases distinguished by the value of $\frac{1}{\psi}$. 
\subsubsection{Case $1<\frac{1}{\psi}<2$}
We treat the expression for $\varphi(\bar{s})$ given in Eq.~(\ref{phi_def}) with a line of steps analogous to those applied above for the saddle-point equation. We employ the Poisson formula to the sum over $k_d$, replace the $r$-summation with an integral, and finally perform the expansion of the Bose function around $\bar{s}=0$. It is here necessary to keep the two leading $\bar{s}$-dependent contributions, so that 
\begin{equation}
g_{\epsi+1}(e^{\bar{s}})- \zeta \Big(\epsi +1 \Big)=\Gamma \Big(-\epsi \Big) |\bar{s}|^{\epsi} -\zeta \Big( \epsi \Big) |\bar{s}|+\dots\;.
\end{equation}
 As a result, in the limit $D\gg\lambda_2$ and for $x \geq 0$ (denoted by superscript $<$), we obtain the following expression for $\omega^<_s$: 
\begin{equation}
\label{omega_s_1}
\frac{\omega_s^<}{k_B T}= -\chi^{d-m} \frac{\Delta^<(x)}{D^{\frac{4}{\psi}-1}}=-\chi^{d-m} \frac{\Delta^<(x)}{D^{2d-m-1}}\;,
\end{equation}
where 
\begin{equation}
\chi=\frac{\lambda_2^2}{\lambda_1}=A\frac{\pi^{3/2}}{2\Gamma(5/4)^2}(t/t_0)^{1/2}
\end{equation}
 is a temperature-independent microscopic length, while  
\begin{align}
\label{Delta_PBC_1}
&\Delta^<(x)= \bigg(\frac{\Gamma(5/4)}{\pi}\bigg)^4\zeta \Big(\epsi \Big)x \sigma(x)^4 \nonumber \\
&+\bigg( \frac{\Gamma(5/4)}{\pi} \bigg)^{\frac{4}{\psi}} \sigma(x)^{\frac{4}{\psi}} \bigg(\Gamma \Big(-\epsi \Big) + \frac{1}{\Gamma(5/4)} \sum_{n=1}^{\infty} F_{\epsi+1}(n \sigma (x)) \bigg) 
\end{align}
represents the scaling function. The quantity $\sigma(x)$ must be determined from the saddle-point equation~(\ref{eqsadle}) as described in the previous sections. For $\frac{1}{\psi}>\frac{5}{4}$ and $x>x_{cr}$ the scaling function is independent of  $x$ and reads
\begin{equation}
\label{delta_<1}
\Delta^<(x)=\frac{4}{\Gamma(5/4)} \Big( \frac{\Gamma(5/4)}{\pi} \Big)^{\frac{4}{\psi}} G \Big(\epsi +1 \Big) \zeta \Big( \frac{4}{\psi} \Big)\;.
\end{equation}
We immediately note that the scaling function is constant for $x>x_{cr}$ and monotonous for $x<x_{cr}$. This implies that it is not an analytical function at $x=x_{cr}$ [we do not exclude however the possibility that it is  smooth (i. e. from the $\mathcal{C}^\infty$ class)]. The corresponding plot is given in Fig.~4 for the physically most relevant case $\frac{1}{\psi}=\frac{5}{4}$.

A few interesting facts are clear from Eq.~(\ref{omega_s_1}). The power law governing the decay of the excess free energy is modified with respect to the standard case: the exponent $\zeta_m=2d-m-1$ replaces the usual value $\zeta_0=d-1$. Such an effect is accompanied by the appearance of a nonuniversal (dimensionful) scale factor $\chi^{d-m}$ multiplying the universal scaling function $\Delta^<(x)$. The result obtained for the exponent $\zeta_m$ agrees with the general prediction of Ref.~\onlinecite{Burgsmuller_2010}, which related $\zeta_m$ to the anisotropy exponent $\theta_A$ via Eq.~(\ref{Diehl_zeta_perp}).
The detailed study of the correlation function of the anisotropic imperfect Bose gas (see Ref.~\onlinecite{Jakubczyk_2018}) shows that $\theta_A=1/2$, which, after plugging into Eq.~(\ref{Diehl_zeta_perp}) yields $\zeta_m=2d-m-1$ in agreement with Eq.~(\ref{omega_s_1}). The profile of the scaling function $\Delta^<(x)$ obtained by solving Eq.~(\ref{eqsadle}) for $\sigma(x)$ and plugging into Eq.~(\ref{Delta_PBC_1}) is plotted in Fig.~4 for the experimentally meaningful case $\frac{1}{\psi}=\frac{5}{4}$ and $x\geq0$. 
\begin{figure}[ht]
\begin{center}
\label{}
\includegraphics[width=7.5cm]{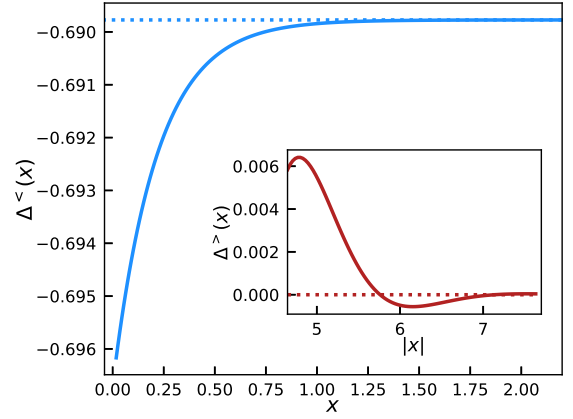}
\caption{The scaling function $\Delta^<(x)$ in the low-temperature phase ($x \geq0$) and $\frac{1}{\psi}=\frac{5}{4}$. The asymptotic values at $x=0$ and $x\to\infty$ correspond to Casimir amplitudes at the transition and in the low-temperature phase, respectively. The negative sign of $\Delta^<(x)$ indicates repulsive character of the Casimir force. The difference between the  values of $\Delta^<(0)$ and $\Delta^<(\infty)$ is tiny. The inset demonstrates the scaling function $\Delta^>(x)$ for $|x|$ sufficiently large so that  $\bar{s}$ may be replaced with its bulk limit. The damped oscillatory behavior reflects the structure of the density-density correlation function\cite{Jakubczyk_2018} and indicates that the sign of the exponentially suppressed interaction depends on the distance $D$.    }
\end{center}
\end{figure} 
The negative sign of $\Delta^<(x)$ indicates repulsive character of the interaction in the low-temperature phase, in clear contrast to the usual situation with periodic boundary conditions. The asymptotic values $\Delta^<(0)$ and $\Delta^<(\infty)$ correspond to Casimir amplitudes at the transition and in the low-temperature phase, respectively. The difference between these two values is rather tiny. 

We now discuss the Casimir-like interaction in the high-temperature phase ($x<0$, denoted by superscript $>$), where the correlation lengths are finite and therefore the effective force is expected to decay exponentially for $D\gg \xi_{\perp, \parallel}$. In this case we obtain an analytical expression for the scaling function in the regime $|x|\gg 1$, where we may replace $\bar{s}$ with its bulk value $s_0$. The  asymptotic behavior of $\omega_s^>$ for $|x|\gg1$ (and $x<0$) is obtained as 
\begin{equation}
\frac{\omega_s^>}{k_B T}= -\chi^{d-m} \frac{\Delta^{>}(x)}{D^{\frac{4}{\psi} -1}}=-\chi^{d-m} \frac{\Delta^>(x)}{D^{2d-m-1}}\;,
\end{equation}
where
\begin{equation}
\label{delta}
\Delta^{>}(x)=\bigg( \frac{\Gamma(5/4)}{\pi} \bigg)^{\frac{4}{\psi}} \sigma(x)^{\frac{4}{\psi}}\sum_{n=1}^{\infty} F_{\epsi+1}(n \sigma (x))\;.
\end{equation}
The scaling function $\Delta^{>}(x)$ displays exponentially damped oscillations deriving from the structure of the function $\phi(k)$ [Eq.~(\ref{phi_function})]. Its profile for $\frac{1}{\psi}=\frac{5}{4}$ is exhibited in the inset of Fig.~4. 

As concerns the dependence of the scaling function ($\Delta^<(x)$ first of all) on dimensionality $\frac{1}{\psi}$, it is clear from Eq.~(\ref{delta_<1}) that its sign and magnitude are controlled by the function $G$, whose sign may change depending on the argument. In fact, $\Delta^<(x)$ features a complex and interesting structure as function of $\frac{1}{\psi}$ resulting in a change of sign of the scaling function (and, in consequence also the Casimir force). This is demonstrated in Fig.~5, where we plot $\Delta^<(0)$ and $\Delta^<(\infty)$ as function of $\frac{1}{\psi}$. Note however that for the physically most meaningful cases (such as $\frac{1}{\psi}=\frac{5}{4}$)  - compare Sec.III) the force is repulsive.  
\begin{figure}[ht]
\begin{center}
\label{}
\includegraphics[width=7.5cm]{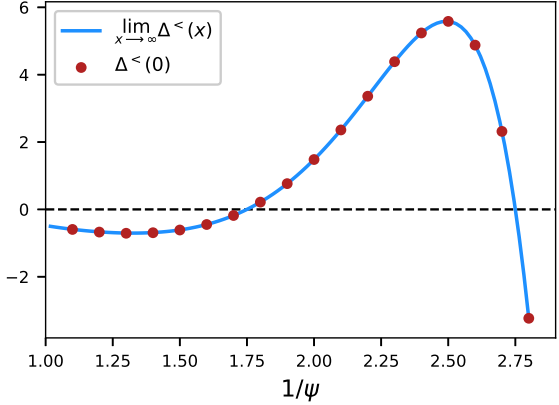}
\caption{The dependence of the Casimir amplitudes $\Delta^<(x=0)$ and $\Delta^<(x\to\infty)$ on $\frac{1}{\psi}$. The difference between the two quantities is nonzero up to $\frac{1}{\psi}=\frac{7}{4}$ (compare Fig.~4), but is not visible in the plot scale. Negative value of $\Delta^<(x=0)$ indicates a repulsive interaction.    }
\end{center}
\end{figure} 
\subsubsection{Case $\frac{1}{\psi}>2$}
The analysis of this case proceeds along the same line as for $1<\frac{1}{\psi}<2$, but significantly simplifies due to vanishing of $\sigma(x)$ obtained from the solution of the saddle-point equation. For fixed $\frac{1}{\psi}$ and $x\geq0$ one obtains a constant scaling function of value given by Eq.~(\ref{delta_<1}). The expression for $\Delta^{>}(x)$ given in Eq.~(\ref{delta}) remains valid also for the present case. The Casimir amplitude is plotted in Fig.~5 together with the results obtained for $\frac{1}{\psi}<2$.
\subsubsection{Casimir force}
Finally, the expression for the Casimir force is obtained via differentiation of the excess free energy. We obtain:
\begin{equation}
\frac{F(D,\mu,T)}{k_B T}=- \chi ^{d-m} \frac{\bar{\Delta}(x)}{D^{2d-m}}
\end{equation}
with
\begin{equation}
\bar{\Delta}(x)=\begin{cases}
        \Big[ (2d-m-1)-(2d-m-4)x \partial_x \Big] \Delta(x), &  1<\epsi<2 \\
        \Big[ (2d-m-1)-4x \partial_x \Big] \Delta(x), &  \epsi >2 .
        \end{cases}
\end{equation}

\subsection{Walls perpendicular to ${\bf{k}}_1$}
We now analyze the complementary situation, where the confining walls are oriented perpendicular to ${\bf{k}}_1$ (i.e. one of the normal directions), and, as we show below, the scaling function for the Casimir energy has completely different properties as compared to the setup discussed above. In essence the computation proceeds along the same line, the difference being that the roles of $k_1$ and $k_d$ are interchanged and Eq.~(\ref{k_d_sum}) becomes replaced by
\begin{equation}
\label{start2}
\sum_{k_1} e^{-r\beta \tilde{\epsilon}_{k_1}}=\frac{D}{\lambda_1\sqrt{r}}+2 \frac{D}{\lambda_1 \sqrt{r}} \sum_{n=1}^{\infty} \exp \bigg(-\pi \frac{D^2 n^2}{\lambda_1^2 r} \bigg)\;,
\end{equation}
where we again used the Poisson summation formula. The saddle point equation is then written as 
\begin{equation}
\begin{aligned}
&\zeta \Big(\frac{1}{\psi}\Big) \bigg(-\frac{\bar{s}}{\mu_c \beta}+\varepsilon\bigg)=g_{\frac{1}{\psi}}(e^{\bar{s}})-\zeta \Big(\epsi\Big)\\
&+2\sum_{r=1}^{\infty}\frac{e^{r \bar{s}} }{r^{\epsi} }\sum _{n=1}^{\infty}\exp \Big(-\pi \frac{D^2 n^2}{\lambda_1^2 r}\Big)
-\frac{\lambda_1^{d-m} \lambda_2^{m}}{V} \sum_{k_1} \frac{1 }{1-e^{\beta \tilde{\epsilon}_{k_1} -\bar{s}}}\;.
\end{aligned}
\end{equation}
For $D\gg \lambda_1$ the $r$-summation can now be transformed into an integral by using the Euler-Maclaurin formula. The resulting integral is expressible via the Bessel function and the saddle-point equation takes the form
\begin{equation}
\label{sad4}
\begin{aligned}
&\zeta \Big(\epsi\Big) \bigg(-\frac{\bar{s}}{\mu_c \beta}+\varepsilon\bigg)=g_{\epsi}(e^{\bar{s}})-\zeta \Big(\epsi\Big)\\
&+\frac{2^{3-\epsi}}{\pi^{\epsi-1}}  \bigg(\frac{\lambda_1}{D} \bigg)^{\frac{2}{\psi}-2} \sum_{n=1}^{\infty}\Big(\frac{\sigma'}{n} \Big)^{\epsi-1} K_{\epsi-1} (n \sigma') 
-\frac{\lambda_1^{d-m} \lambda_2^{m}}{V} \sum_{k_1} \frac{1 }{1-e^{\beta \tilde{\epsilon}_{k_1} -\bar{s}}}\;,
\end{aligned}
\end{equation}
where 
\begin{equation} 
\sigma'=2\sqrt{\pi}\frac{D}{\lambda_1} |\bar{s}|^{1/2}\;.
\end{equation}
Expanding the Bose function for $|\bar{s}|\ll1$ we again encounter the different cases depending on the value of $\frac{1}{\psi}$. Introducing
\begin{equation}
x'=\begin{cases}
        \varepsilon \Big(\frac{D}{\lambda_1} \Big)^{\frac{2}{\psi}-2}, &  1<\epsi<2 \\
        \varepsilon \Big(\frac{D}{\lambda_1} \Big)^2, &  \epsi >2
        \end{cases}
\end{equation}
the saddle-point equation is written as 
\begin{equation}
\begin{aligned}
\zeta \Big(\epsi\Big) x' \pi^{\epsi-1}&=\frac{\Gamma(1-\epsi)}{2^{\frac{2}{\psi}-2}} \sigma'^{\frac{2}{\psi}-2} \\
&+2^{3-\epsi} \sigma'^{\epsi-1}\sum_{n=1}^{\infty} (n^{-1})^{\epsi-1} K_{\epsi-1} (n \sigma') \\
&+\pi^{\epsi-1} \Big(\frac{D}{\lambda_1} \Big)^{\frac{2}{\psi}-2} \frac{\lambda_1^{d-m} \lambda_2^{m}}{V} \sum_{k_1} \frac{1 }{e^{\beta \tilde{\epsilon}_{k_1} -\bar{s}}-1} 
\label{speee}
\end{aligned}
\end{equation}
for $1<\frac{1}{\psi}<2$ and 
\begin{equation}
\begin{aligned}
\zeta \Big(\epsi \Big) x' &= -\frac{1}{4\pi} \sigma'^2 \bigg(\zeta\big(\epsi-1\big)+\frac{1}{\mu_c \beta} \bigg)\\
&+ \frac{2^{3-\epsi}}{\pi^{\epsi-1}} \Big( \frac{\lambda_1}{D} \Big)^{\frac{2}{\psi}-4 } \sigma'^{\epsi-1}\sum_{n=1}^{\infty} (n^{-1})^{\epsi-1} K_{\epsi-1} (n \sigma') +\\
&+\Big(\frac{D}{\lambda_1} \Big)^{2} \frac{\lambda_1^{d-m} \lambda_2^{m}}{V} \sum_{k_1} \frac{1 }{e^{\beta \tilde{\epsilon}_{k_1} -\bar{s}}-1}
\end{aligned}
\end{equation}
for $\frac{1}{\psi}>2$, providing a relation between the quantities $x'$ and $\sigma'$, which is necessary to evaluate the excess free energy via Eq.~(\ref{excess}). For $\epsi=2$ again logarithmic corrections appear. Focusing now on the case $1<\frac{1}{\psi}<2$ we note that the last term on the RHS of Eq.~(\ref{speee}) can be neglected as long as $\sigma'$ (which solves Eq.~(\ref{speee})) is finite. For $\frac{1}{\psi}>\frac{3}{2}$ the RHS is in such a case bounded from above by 
\begin{equation}
\lim_{\sigma'\to 0}2^{3-\frac{1}{\psi}}\sigma'^{\frac{1}{\psi}-1}\sum_{n=1}^{\infty}n^{1-\frac{1}{\psi}}K_{\frac{1}{\psi}-1}(n\sigma')=2\Gamma\left(\frac{1}{\psi}-1\right)\zeta\left(\frac{2}{\psi}-2\right)\;.
\end{equation}
This implies that for $x'>x'_{cr}>0$, where
\begin{equation}
x'_{cr} \Big( \frac{1}{\psi} \Big)=\frac{2}{\zeta(\frac{1}{\psi})\pi^{\frac{1}{\psi}-1}}\Gamma\left(\frac{1}{\psi}-1\right)\zeta\left(\frac{2}{\psi}-2\right)
\end{equation} 
$\sigma'$ vanishes. In consequence, the last term in Eq.~(\ref{speee}) cannot be neglected. As a result, one obtains $\sigma'=0$ for $x'>x'_{cr} (\frac{1}{\psi})$  (and $\frac{3}{2}<\frac{1}{\psi}<2$). The scaling function is then constant (see below). For $\frac{1}{\psi}>2$ one obtains $\sigma'=0$ for any value of $x'>0$, whereas for $x'\leq0$ we find
\begin{equation}
\sigma'(x')= \Bigg( \frac{4 \pi|x'|}{\frac{1}{\mu_c \beta} +\zeta(\epsi-1)/ \zeta(\epsi)} \Bigg)^{1/2}.
\end{equation}
As concerns the excess free energy, for the present case one finds 
\begin{equation}
\label{vphi}
\begin{aligned}
\varphi(\bar{s})&=-\frac{\bar{s}^2}{2 a \beta}-|\bar{s}|\frac{\mu}{a}-\frac{1}{\lambda_1^{d-m} \lambda_2^m}g_{\epsi+1}(e^{\bar{s}})\\
&-\frac{1}{\lambda_1^{d-m} \lambda_2^m} \frac{2^{2-\epsi}}{\pi^{\epsi}} \Big( \frac{\lambda_1}{D} \Big)^{\frac{2}{\psi} } \sigma'^{\epsi}\sum_{n=1}^{\infty} (n^{-1})^{\epsi} K_{\epsi} (n \sigma')+\\
&+\frac{1}{V}\sum_{k_1} \log \Big( 1-e^{\bar{s}-\beta \tilde{\epsilon}_{k_1}} \Big)\;,
\end{aligned}
\end{equation}
and the last term always gives a vanishing contribution for $L\to\infty$. For $x'\geq 0$ Eq.~(\ref{excess}) can now be cast in the form 
\begin{equation}
\label{omegaperp}
\frac{\omega_s^<}{k_B T}=-\frac{1}{\chi^{m/2}} \frac{\Delta^{'<}(x')}{D^{\frac{2}{\psi}-1}}=-\frac{1}{\chi^{m/2}} \frac{\Delta^{'<}(x')}{D^{d-m/2-1}}
\end{equation}
where  the scaling function is given by
\begin{equation}
\begin{aligned}
\Delta^{'<}(x')&=\frac{\zeta(\epsi)}{4 \pi} \sigma' (x')^2 \, x'+ \frac{\Gamma(-\epsi)}{2^{\frac{2}{\psi}} \pi^{\epsi}} \sigma'(x')^{\frac{2}{\psi}}\\
&+\frac{2^{2-\epsi}}{\pi^{\epsi}}  \sigma'(x')^{\epsi}\sum_{n=1}^{\infty} (n^{-1})^{\epsi} K_{\epsi} (n \sigma'(x'))
\end{aligned}
\end{equation}
and the relation $\sigma'(x')$ is determined from the solution of the saddle-point equation. For $\sigma'=0$ the scaling function is constant and takes the value $2 \zeta (\frac{2}{\psi}) \Gamma(\epsi)/\pi^{\epsi}$. In the high-temperature phase ($x<0$) we obtain the following expression for the scaling function for large $|x|$:
\begin{equation}
\Delta_{}^{'>}(x')=\frac{2^{2-\epsi}}{\pi^{\epsi}}  \sigma'(x')^{\epsi}\sum_{n=1}^{\infty} (n^{-1})^{\epsi} K_{\epsi} (n \sigma'(x')).
\end{equation}
Finally, the Casimir force is given by
\begin{equation}
\frac{F(D,\mu,T)}{k_b T}=-\frac{1}{\chi ^{m/2}} \frac{\bar{\Delta'}(x')}{D^{d-m/2}}
\end{equation}
with
\begin{equation}
\bar{\Delta'}(x')=\begin{cases}
        \Big[ (d-\frac{m}{2}-1)-(d-\frac{m}{2}-2)x' \partial_{x'} \Big] \Delta(x'), &  1<\epsi<2 \\
        \Big[ (d-\frac{m}{2}-1)-2x' \partial_{x'} \Big] \Delta'(x'), &  \epsi >2 .
        \end{cases}
\end{equation}
We conclude that also for the present situation where the confining walls are oriented perpendicular to ${\bf{k}}_1$ the power law governing the decay of the Casimir interaction is modified, which is accompanied by the appearance of a nonuniversal, dimensionful scale factor $\frac{1}{\chi^{m/2}}$. The obtained exponent $\zeta_m'=d-\frac{m}{2}-1$ again agrees with the form $\zeta_m'=d-m(1-\theta_A)-1$ predicted in Ref.~\onlinecite{Burgsmuller_2010} [compare Eq.~(\ref{Diehl_zeta_par})]. The scaling function is monotonous and positive in each of the phases and points at attractive interaction, in contrast to the previous case of walls oriented perpendicular to ${\bf{k}}_d$. The oscillations in the high-temperature phase are also absent. The profile of $\Delta^{'<}(x')$ is identical to that derived in Ref.~(\onlinecite{Napiorkowski_2013}) for isotropic continuum  case upon identifying $\frac{1}{\psi}\to \frac{d}{2}$ so that the net effect of the anisotropy is the modification of the decay exponent $\zeta_m'$ accompanied by the appearance of the scale factor $\frac{1}{\chi^{m/2}}$ [see Eq.~(\ref{omegaperp})]. This holds true also for $\frac{1}{\psi}>2$ as well as  $1<\frac{1}{\psi}<2$ and $x'<0$. We note however, that the existence of $x'_{cr}(\frac{1}{\psi})$ and the fact that the scaling function is constant for $x'>x'_{cr}(\frac{1}{\psi})$  for $\frac{3}{2}<\frac{1}{\psi}<2$ was not discussed in the study of Ref.~(\onlinecite{Napiorkowski_2013}). This result implies that the scaling function $\Delta^{'<}(x')$ is not analytical at $x'_{cr}(\frac{1}{\psi})$. This feature is related to the phase transition occurring at finite $D$ and should be shared by the entire $O(N\to\infty)$ universality class (also for the isotropic case) in dimensionality $3<d<4$.\cite{Dantchev_2004} We once again invoke here the recently established fact\cite{Diehl_2017} that the critical behavior of the  imperfect Bose gas with periodic boundary conditions maps exactly onto the corresponding classical $O(2N)$ model in the limit $N\to\infty$. The scaling functions of these two models are the same modulo a global factor of two.\cite{Dantchev_1996, Napiorkowski_2013, Diehl_2017}  

\section{Summary and outlook}
In this paper we addressed the Bose gas in optical lattices focusing on effects induced by spatial anisotropies which may be controlled by varying the lattice parameters. By suitably tuning the couplings, the system is driven into a strongly anisotropic setup, where condensation is characterized by two divergent length scales ($\xi_{\parallel}$ and $\xi_\perp$) related by the anisotropy exponent $\theta_A$ (so that $\xi_\perp\sim\xi_{\parallel}^{\theta_A}$) and by fractional effective spatial dimensions. We addressed two aspects of the system induced by such anisotropies, the first one being related to dimensional crossovers in the bulk Bose-Einstein condensation (Sec.~IV), the other to Casimir interactions (Sec.~V). We employed the imperfect Bose gas as the prototypical model. As we argued at the end of Sec.~IV, we believe our major conclusions concerning the dimensional crossovers in the bulk should not be restricted to mean-field models. Indeed, the imperfect Bose gas is known to be closely related to the $O(N\to\infty)$ universality class, while realistic condensation corresponds to $N=2$. It is however known that at least some of the features studied here (the universal asymptotics of the $T_c$-line in particular) are insensitive to the symmetry-breaking involved. Dimensional crossovers and the idea of tuning the system across the phase transition by varying dimensionality seems to be a problem of current experimental interest\cite{Ilg_2018, Noh_2020} and we have provided an analytical understanding of these effects realized by tuning the lattice hoppings. Particularly interesting situations arise when one of the involved effective dimensionalities is located at or below the lower critical dimension ($d_{eff}=2$) for condensation. The hoppings may then be tuned to completely expel the condensate out of the system. We clarified the mechanism that governs this behavior. 

Our results for the scaling function of the Casimir energy (Sec.~V) indicate a rather unusual behavior presumably generic for systems characterized by dispersions varying as $\sim k^4$ (which however calls for further studies). Starting from a microscopic level and performing an exact analysis we have confirmed the picture of Ref.~\onlinecite{Burgsmuller_2010} concerning modifications of the decay exponent for the Casimir interaction. This is necessarily accompanied by the appearance of a nonuniversal dimensionful scale factor. We focused on periodic boundary conditions and addressed  two configurations of the confining walls. In the first case, the walls are perpendicular to a direction characterized by a $\sim k^4$ dispersion; in the other setup the dispersion in the perpendicular direction is of the type $\sim k^2$. In the former situation and at physically most relevant effective dimensionalities, the obtained Casimir interaction turns out to be repulsive below and at the critical temperature $T\leq T_c$. In this regime we evaluated the entire profile of the universal scaling function, which turns out to be  monotonous for positive values of the scaling variable $x$ (for $T\leq T_c$). In contrast, for $x<0$ the scaling function shows exponentially damped oscillatory behavior and changes sign upon varying the distance $D$. In the present setup these effects are encoded in the rich structure of the function $\phi(k)$ (Fourier transform of the quartic Gaussian). By virtue of universality we expect similar behavior to  apply to the entire $O(N\to\infty)$ universality class (up to proportionality factors understood in Ref.~\onlinecite{Diehl_2017} for the isotropic case). An extension to finite $N$ is not an easy enterprise as is clear from Ref.~\onlinecite{Burgsmuller_2010}. For the standard case of isotropic $O(N)$ models the profile of the scaling function (but usually not its sign) may be different as compared to their $N\to\infty$ limiting shapes.\cite{Dantchev_1996, Vasilyev_2009} 

The possibility of obtaining repulsive Casimir forces was recently studied in a number of contexts.\cite{Dohm_2008, Dohm_2011, Bimonte_2011, Jakubczyk_2016_2, Rajabpour_2016, Sadhukhan_2017, Flachi_2017, Faruk_2018, Dohm_2018, Voronina_2019, Voronina_2019_2} In many situations, such scenarios are realized by varying the boundary conditions, which, to some extent may also be controlled experimentally by engineering the surface properties (see e.g. \onlinecite{Soyka_2008, Nellen_2009, Trondle_2011}). This aspect was also addressed by means of numerical simulations (see e.g.~\onlinecite{Vasilyev_2009, Vasilyev_2007, Vasilyev_2013, Toldin_2013, Toldin_2015, Toldin_2015_2} ). 
We point out, however, that the possibility of obtaining repulsive Casimir interaction for periodic boundary conditions is quite uncommon. Further theoretical verification of this possibility (at finite $N$ in particular) is an interesting direction for future studies. 

As concerns the situation with $\sim k^2$ dispersion in the direction perpendicular to the confining walls, we have identified a similar effect of modifying the decay exponent for the Casimir energy, which is necessarily accompanied by emergence of a nonuniversal, dimensionful scale factor in the expression for the excess free energy. Once this is factored out, one recovers the universal scaling function identical to that obtained for the isotropic case, however in lower dimensionality $d_{eff}<3$. The resulting Casimir interaction is then always attractive and shows no oscillations of the type observed in the high-temperature phase in the case where the dispersion in the direction perpendicular to the  walls is of the $\sim k^4$ type. For $T>T_c$ the Casimir force is always exponentially suppressed and its form (monotonous or oscillating) reflects the properties of the correlation function in the direction perpendicular to the confining walls. A classical case, where similar two situations should be possible to achieve is provided by classical supercritical fluids, where the pair correlation function changes its behavior across the so-called Widom-Fisher line\cite{Fisher_1969} despite the absence of a phase transition.   

On the experimental side, besides the present context, anisotropic scale invariance is also present at the Lifshitz points as well as in liquid crystals, to which (by virtue of universality) our results concerning the Casimir force may perhaps also apply. Realization of our model would require a lattice with anisotropic, long-range (next-nearest-neighbours at least) hopping parameters. Three-dimensional optical lattices with direction-dependend hoppings were reported in \cite{Imriska_2014, Greif_2013}. Additionally, let us note that it is possible to simulate higher dimensionalities using optical lattice setups \cite{Boada_2012, Boada_2015}. While the dimensional crossover as analyzed in the present study, might be within range of current (or near future) experimental techniques on ultracold gases in optical lattices, accurate observation of the Casimir effect in such systems might appear harder to achieve. In this aspect, classical systems such as liquid crystals appear perhaps as more promising candidates. Note that, due to the presence of Goldstone modes, long-ranged Casimir forces in such systems occur in the entire low-$T$ phase, so that no tuning to the critical point is required. A verification of our predictions might be accomplished by inspecting very crude features of the Casimir effect: for example the dependence of the Casimir force sign on the orientation of the confining walls. In addition to experiments, our predictions are certainly also open to verification via numerical simulations.
\begin{acknowledgments}
We are grateful to Hans Werner Diehl, Marek Napi\'orkowski and  Piotr Nowakowski for discussions as well as reading the manuscript and providing helpful suggestions. We also thank Daniel Dantchev for a useful correspondence. PJ acknowledges support from the Polish National Science Center via  2017/26/E/ST3/00211. 
\end{acknowledgments}

\bibliography{/Users/pjak/Desktop/refs}{} 
\bibliographystyle{apsrev4-1}

\end{document}